\pdfoutput=1
\documentclass{llncs}
\usepackage{makeidx} 
\usepackage{multirow}

\usepackage{graphicx}
\usepackage{csquotes}
\usepackage{url}
\ifundef\doi
  {\DeclareUrlCommand\doi{}}
  {}
\usepackage{subfig}
\usepackage{amsmath}
\usepackage{amssymb}
\DeclareMathOperator*{\argmin}{argmin}

\begin{document}
%
%
\pagestyle{headings}  
%

\title{3D Deep Affine-Invariant Shape Learning for Brain MR Image Segmentation}
\titlerunning{Shape Learning}  
%
\author{Zhou He \and Siqi Bao \and 
Albert Chung}
\institute{Department of Computer Science and Engineering \newline The Hong Kong University of Science and Technology}

\maketitle              
\begin{abstract}
Recent advancements in medical image segmentation techniques have achieved compelling results. However, most of the widely used approaches do not take into account any prior knowledge about the shape of the biomedical structures being segmented. More recently, some works have presented approaches to incorporate shape information. However, many of them are indeed introducing more parameters to the segmentation network to learn the general features, which any segmentation network is able learn, instead of specifically \textit{shape} features. In this paper, we present a novel approach that seamlessly integrates the shape information into the segmentation network. Experiments on human brain MRI segmentation demonstrate that our approach can achieve a lower Hausdorff distance and higher Dice coefficient than the state-of-the-art approaches.

\end{abstract}

\section{Introduction}
A variety of approaches have been adopted to address the challenging problem of 3D medical image segmentation, such as 3D U-Net \cite{3dunet} and V-Net \cite{vnet}, which have been proven to be highly effective. These approaches, however, simply transplant the 2D image semantic segmentation algorithms to a 3D medical image analysis context. They have little awareness to the fact that 3D medical structures of the same class, unlike objects in 2D natural images, in general have similar shapes. For example, for a 2D natural image segmentation task on the class of 'person', different persons could be very different in shape since a person may have different poses when being photographed, e.g., arms opened/closed, sitting/standing, etc. For the segmentation on biomedical structures such as human caudate nucleus, all caudate nuclei have very similar shape with little structural variation. However, this information is rarely used in deep learning-based 3D medical image segmentation. While some recent literature has introduced some approaches to leverage shape information, many of them are merely introducing more hyperparameters to the network to increase its capacity, while not actually using exactly the \textit{shape} information. \par
In this paper, we present a novel approach which incorporates the information about the shape of the segmentation target into the loss function of a general 3D segmentation network. This shape information is deep-learned from a fully convolutional network, whose feature map of the final layer (defined as the \textbf{shape signature}) captures the important global shape information. We first pre-train this shape-learning network by ground truth label maps that have undergone different affine transformations, and then have the weights of this network fixed. This shape-learning network will then be able to capture the essential shape information that is invariant to affine transformation. Afterwards, when training the segmentation network, the prediction label map and ground truth label map will both be fed into the pre-trained shape-learning network, and the Euclidean distance between their shape signatures will quantify the dissimilarity in shape between the segmentation prediction and ground truth. This shape loss is then added to the loss function of the segmentation network to facilitate the training. \par
Our main contributions are summarized as follows:
\begin{enumerate}
    \item Designed a novel shape-learning network that is able to capture the affine-invariant global shape information in the final feature map;
    \item Incorporated the shape dissimilarity information to the segmentation network, making it shape-aware;
\end{enumerate}

\section{Related Work}
We start by reviewing related prior works on general medical image segmentation, and the utilization of shape information.
\subsection{Medical Image Segmentation}
Deep learning-based image semantic segmentation became highly successful since the emergence of Fully-convolutional Network (FCN) \cite{fcn}. This approach has later been adapted to a biomedical image segmentation setting with the novel design of U-Net \cite{unet}, which contains skip connections between the contracting path and expanding path so that the intricate details in biomedical images can be kept. Recently, U-Net has been modified to accommodate 3D volumes by replacing all the 2D convolutions and convolution transposes by their 3D counterparts, as described in 3D U-Net \cite{3dunet}. Apart from the change in network architecture, some other adaptations have been made to make CNNs more compatible with medical image segmentation. For instance, in V-Net \cite{vnet}, the loss function is derived from Dice coefficient which is a common metric in medical image segmentation.

\subsection{The Utilization of Shape Information in Segmentation}
Some prior works claimed to have leveraged the shape information of biomedical structures for segmentation purpose. \cite{shapeprior} introduced an autoencoder known as Shape Regularization Network (SRN) that regularizes the segmentation result to make it conform to the shape it should have. Its functions include eliminating any noisy part from the general shape, or filling up any holes in the preliminary segmentation result. A more recent work Anatomically Constrained Neural Networks (ACNN) \cite{acnn} used an autoencoder to learn the shape by training that autoencoder to reconstruct a label map itself, and used the Euclidean distance between the bottleneck layers of the autoencoder to quantify the dissimilarity in shape.\par

A commonality among these prior works is that they introduce another network which is trained to capture the shape information, and this network is then used to guide the segmentation network. However, the shape learner in SRN and ACNN are both learning the general features of a 3D structure, including position, volume, shape, etc, instead of specifically learning the \textit{shape}. Inspired by these issues, we propose an approach to learn the essential shape information that is invariant to affine transformations.

\section{Methodology}
\subsection{Overview}
While the term \textit{shape} may have many definitions in different settings, in the medical image segmentation setting here, we define it to be the intrinsic properties of the 3D biomedical structures that are invariant to spatial affine transformations, including rotation, translation and scaling, etc. The network architecture used in our approach is composed of two parts, where the first part is to capture the shape information, and the second part to use it.  \par
\begin{figure}
\centering
\includegraphics[width=0.83\textwidth]{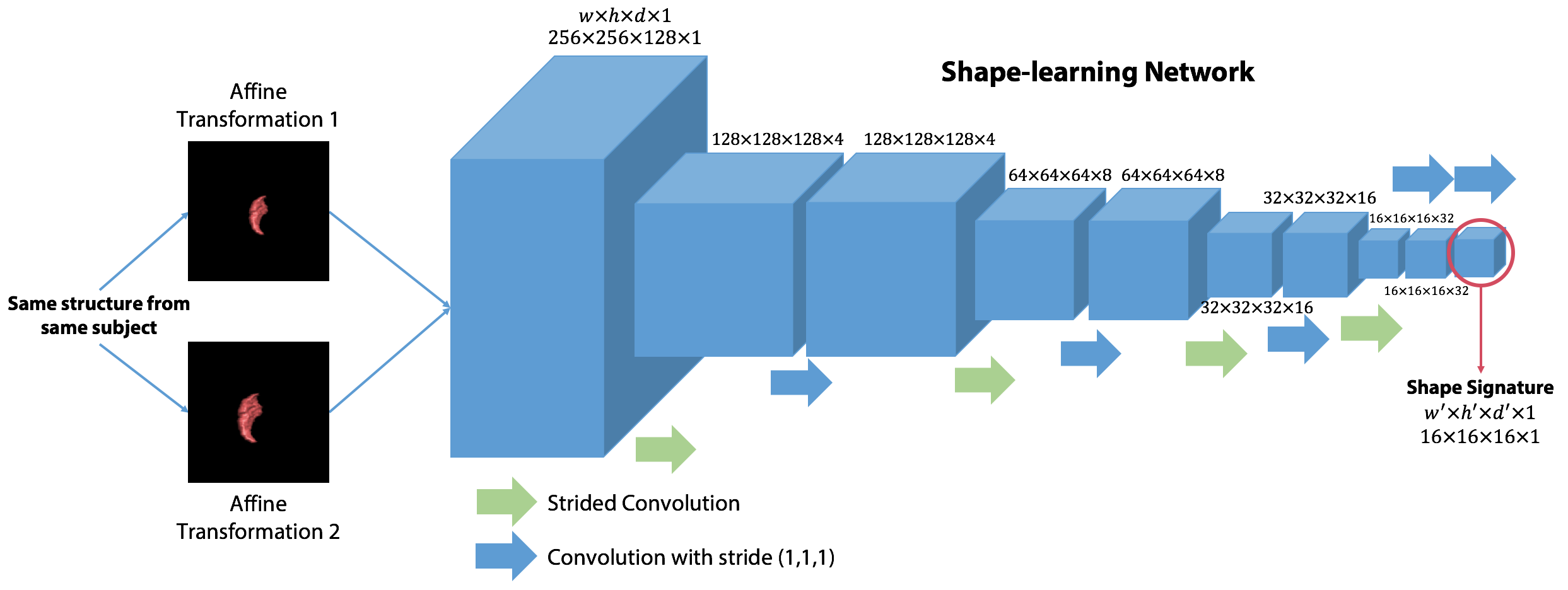}
\caption{Architecture of the shape-learning network.}
\label{cae}
\end{figure}
\begin{figure}
\centering
\includegraphics[width=0.93\textwidth]{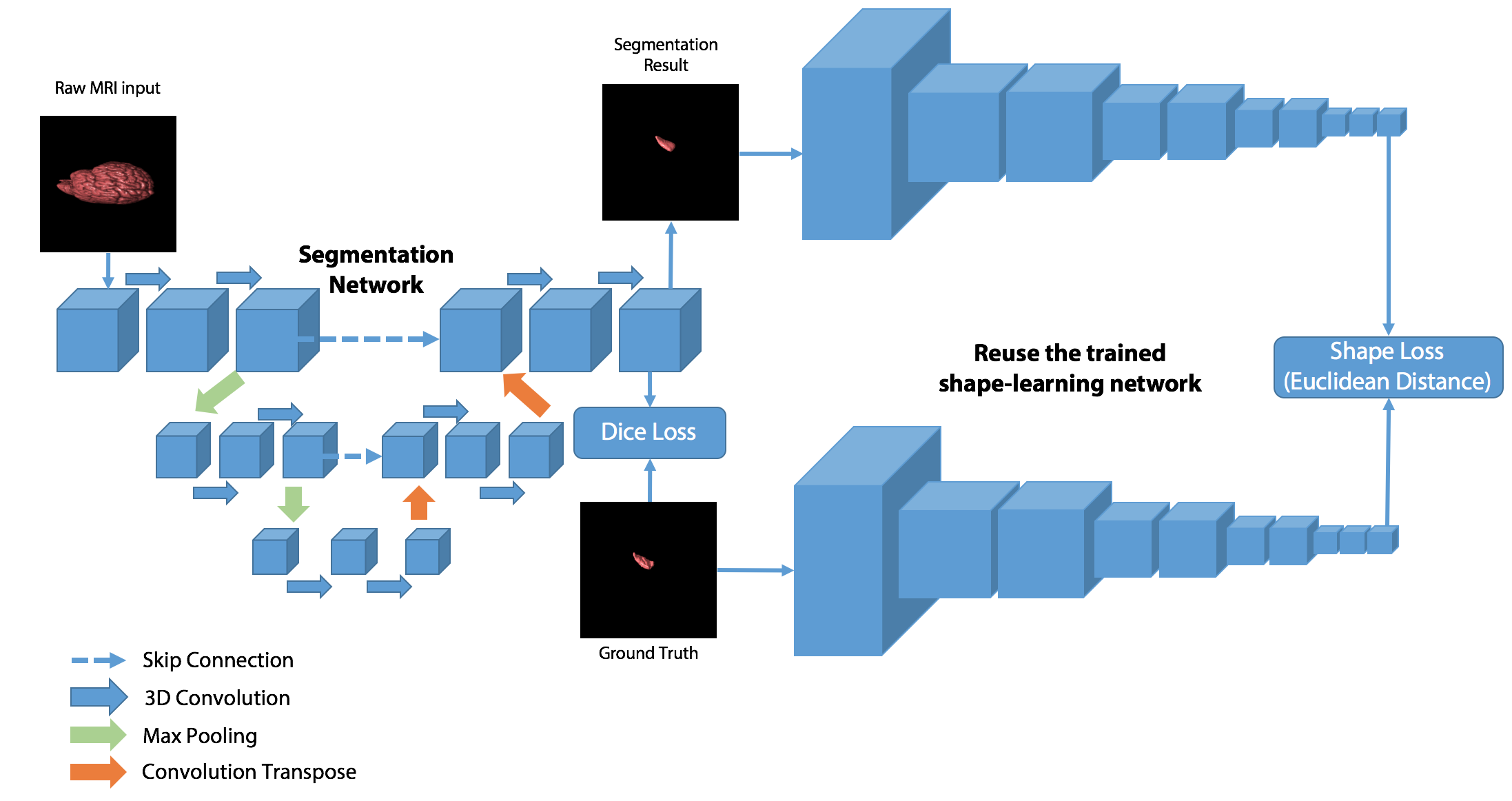}
\caption{Diagram of the full network architecture. The segmentation network, following a 3D U-Net architecture, is shown on the left, and the pre-trained shape-learning network that extracts the shape signature is shown on the right. Number of channels is not reflected on this diagram for brevity.}
\label{architecture}
\end{figure}

Concretely, the first part, defined as \textbf{shape-learning network}, is a 3D fully convolutional neural network (ConvNet), with the input being the raw binary 3D label map of the biomedical structure being segmented, and the output being a one-channel low-resolution feature map (hereafter referred to as \textbf{shape signature}). The second part, defined as \textbf{shape-guided segmentation network}, is a segmentation network with the architecture modeled after the 3D U-Net \cite{3dunet} and loss function being the sum of Dice loss and shape loss. The role of the shape-learning network is to learn the \textbf{shape signature} of a 3D binary label map, and this information would later become a part of the loss function of the segmentation network. The architecture of the shape-learning network is shown in Figure \ref{cae}, while the complete illustration showing the full network architecture is attached at Figure \ref{architecture}.

\subsection{Shape-learning Network}
The first step is to train the shape-learning network. In every iteration of training, we feed the shape-learning network with two binary label maps which are the \textbf{same structure} from the \textbf{same subject} that have gone through \textbf{different affine transformations}. Since these two label maps come from the same subject's same structure under different affine transformations, we call them an \textit{affine pair}, and argue that they contain exactly the \textbf{same shape information}. If the network was able to capture shape information well, the difference between the shape signatures of these two label maps from the same \textit{affine pair} should be small. We therefore compute the Euclidean distance between the shape signatures of these two label maps, and use this difference in Euclidean distance as loss and propagate the loss through the entire network and to update the network weights.\par
Let the ground truth label map be $M \in \mathbb{R}^{w\times h\times d}$ and the shape signature be $\Hat{M}\in \mathbb{R}^{w'\times h'\times d'}$ where $w', h', d'$ are much smaller than $w, h, d$ respectively, the shape-learning network is essentially a non-linear mapping from $M$ to $\Hat{M}$, namely $\Hat{M}=g_{\theta}(M)$, where $\theta$ is the weights in the convolutional layers of this network. Given this shape-learning network $g_\theta$, the shape loss between two binary label maps $M_1, M_2 \in \mathbb{R}^{w\times h\times d}$ is therefore $$\mathcal{L}_{shape}(M_1,M_2)=\|g_{\theta}(M_1)-g_{\theta}(M_2)\|_2$$
Training this shape-learning network therefore essentially means finding the $\theta$ that satisfies $$\theta = \argmin_{\theta} \mathcal{L}_{shape}(M_1,M_2)$$ where $M_1$ and $M_2$ are two instances of the same structure in the same subject, that have gone through different random affine transformation. After the training is finished, given a label map $M$, $g_{\theta}(M)$ gives the shape signature of this label map.

\subsection{Shape-guided Segmentation Network}

After the training of the shape-learning network is finished, we then train the segmentation network which is responsible for generating the segmentation label map. The segmentation network is a mapping $f$ from the input (the raw voxels of brain MR image) $I$ to the segmentation result $\Tilde{M}$ defined as $\Tilde{M}=f_{W}(I)$ where $W$ is the weights of the segmentation network. The difference between the segmentation result $\Tilde{M}$ and ground truth label map $M$ is first measured by the Dice loss defined as$$\mathcal{L}_{dice}(M,\Tilde{M})=1-\frac{2\sum_{i}{M_i}{\Tilde{M_i}}}{\sum_{i}{M_i}+\sum_{i}{\Tilde{M_i}}}$$ And by the definition given in the previous section, the shape loss between $\Tilde{M}$ and the ground truth $M$ is defined as
$$\mathcal{L}_{shape}(M,\Tilde{M})=\|  g_{\theta}(M) - g_{\theta}(\Tilde{M}) \|_2$$  After adding the shape loss term to the Dice loss, we obtain the total loss which is  $$\mathcal{L}_{total}(M,\Tilde{M}) = \mathcal{L}_{dice}(M,\Tilde{M}) + \alpha \mathcal{L}_{shape}(M,\Tilde{M})$$ where $\alpha$ is a hyperparameter that balances the weights of Dice loss and shape loss. The weights $W$ of the segmentation network is $$W=\argmin_{W}\mathcal{L}_{total}(M,\Tilde{M})$$ and the segmentation network can be trained by the Stochastic Gradient Descent (SGD) with backpropagation since the entire pipeline is differentiable end-to-end.

\section{Experiments}
\subsection{Experimental Setup}
Experiments have been implemented on the human left and right caudate nucleus , as well as left and right hippocampus in the LONI Probabilistic Brain Atlas (LPBA40) dataset \cite{lpba}, which is a publicly available series of maps of human brain anatomic regions. The Magnetic Resonance (MR) images in the native space are used as raw input, while the label maps in the delineation space are the ground truth labels.\par
All MRI inputs and their corresponding labels are preprocessed and cropped to a region of 256$\times$256$\times$128 in size, which is identical for every subject. Raw MRI inputs are preprocessed so that the original 12-bit image representation is normalized to a mean of 0.0 and standard deviation of 1.0. The label maps are further preprocessed for left and right caudate nuclei respectively, so that the label maps of both structures are binary three-dimensional arrays. Data augmentation operations on the training data include randomly rotating the object in 3D space up to 8 degrees, randomly scaling the object from 0.85 times to 1.15 times, as well as randomly translating the object. Left and right caudate nucleus and left and right hippocampus are all processed separately and are ran in separate experiments. Note that these transformations are also used in preparing an affine pair when training the shape-learning network, which requires the same structure to go through two random affine transformations.\par
\subsection{Training the Shape-learning Network}

\begin{table}[htp]
\begin{center}
\begin{tabular}{|c| c| c|}
 \hline
  & Same subject & Different subjects \\
 \hline
 Left Caudate & \textbf{0.120} & 0.316 \\
 \hline
  Right Caudate & \textbf{0.083} & 0.210 \\
 \hline 
 Left Hippocampus & \textbf{0.251} & 0.787 \\
 \hline
  Right Hippocampus & \textbf{0.088} & 0.277\\
 \hline
\end{tabular}
\vspace{5mm}
\caption{Average shape loss of 50 random \textit{affine pairs} of the four biomedical structures tested, when the pairs are drawn from the same subject or a different subject.}
\label{shapelearnerperf}
\end{center}
\end{table}
\vspace{-10mm}

We first train the shape-learning network, and demonstrate why it is able to capture the essential shape information in the shape signature layer. The shape-learning network was trained with \textit{affine pairs}, where the binary label maps have both gone through a random affine transformation that was employed in the data augmentation step. On each structure, we train for 200 iterations with batch size 1 and learning rate $1\times10^{-4}$ on Adam Optimizer \cite{adam}. Experimental results illustrated in Table \ref{shapelearnerperf} demonstrate that the average difference in shape signature between affine pairs of the same subject's same structure is much lower than the average shape difference between pairs from different subjects. Therefore, a well-trained shape-learning network is able to capture a structure's essential shape information that is invariant to affine transformations.\par 

\subsection{Training the Segmentation Network}
\begin{figure}
\centering
\includegraphics[width=0.63\textwidth]{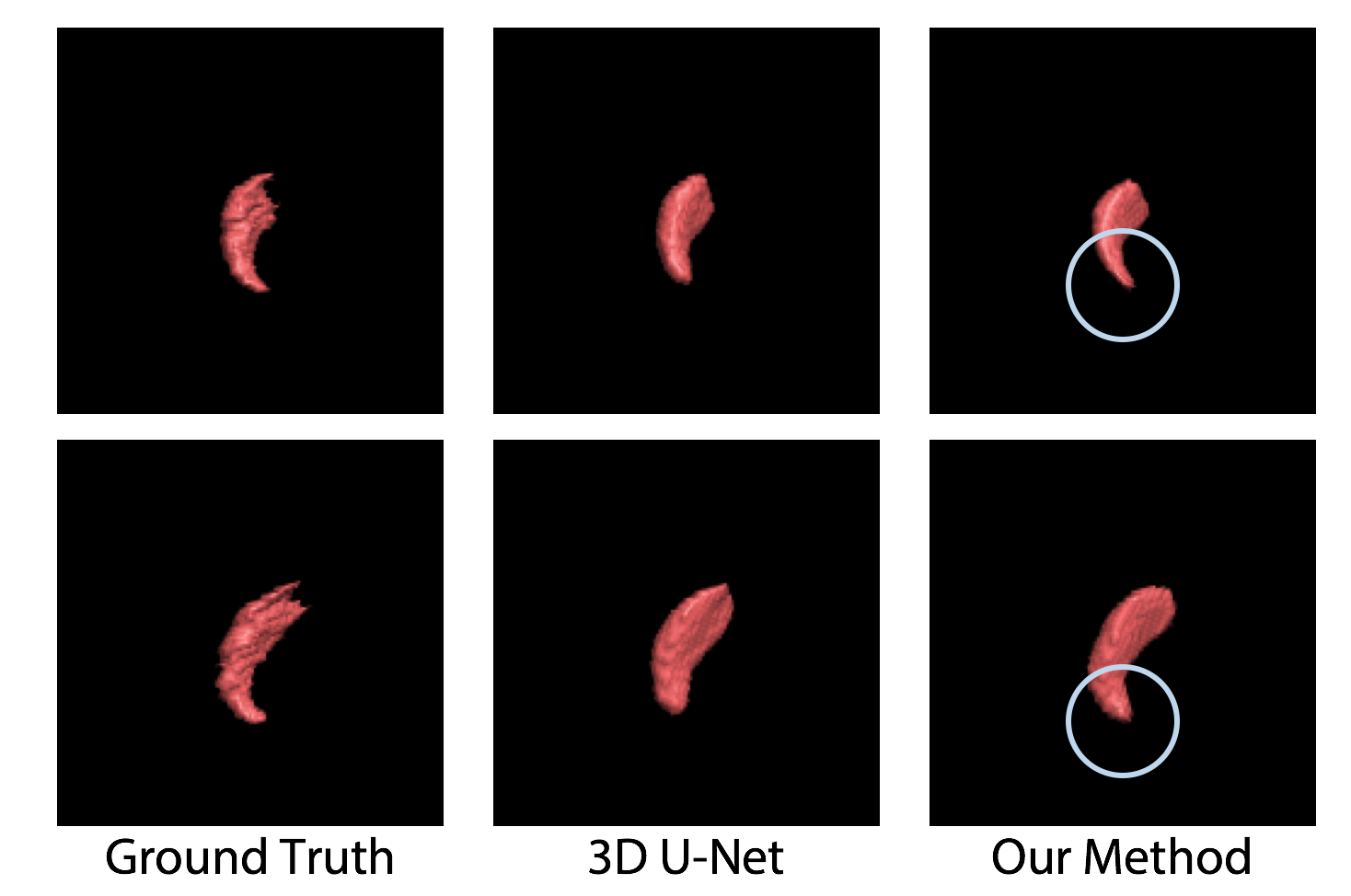}
\caption{Visual Results of our approach compared with 3D U-Net.}
\label{segresult}
\end{figure}

After finish training the shape-learning network, we freeze its weights and train the segmentation network. The experiments were also run with a batch size of 1, with the optimizer being Adam Optimizer \cite{adam} and the learning rate being $1\times10^{-4}$. The weight of shape loss $\alpha$ was chosen experimentally to be $0.1$, and the models of left and right caudate nucleus and left and right hippocampus are first trained without shape loss for 800 iterations, and then trained with shape loss for another 400 iterations. To prevent the shape loss term from being extremely large, we experimentally set it to be capped at 1.0. As ablation experiments, we also run experiments with the same set of hyperparameters and the same dataset with a 3D U-Net model as a comparison. Note that the 3D U-Net here refers to the U-shape network in \cite{3dunet} trained with only Dice loss. Experimental results of Dice coefficient and Hausdorff distance on left and right caudate nucleus of 3D U-Net and our method are listed in Table \ref{lcn}, while the visual results are shown in Figure \ref{segresult}.

\begin{table}[htp]
\begin{center}
\begin{tabular}{|c|c| c| c|} 
 \hline
 Structure & Metric & 3D U-Net & Our method \\
 \hline
 \multirow{2}{*}{Left Caudate} & Dice & 0.831 & \textbf{0.835} \\
 \cline{2-4} & HD & 5.472 & \textbf{5.299} \\ 
 \hline
 
  \multirow{2}{*}{Right Caudate} & Dice & 0.782 & \textbf{0.820} \\
 \cline{2-4} & HD & 6.369 & \textbf{5.004} \\ 
 \hline
 
  \multirow{2}{*}{Left Hippocampus} & Dice & 0.771 & \textbf{0.793} \\
 \cline{2-4} & HD & 20.170 & \textbf{5.843} \\ 
 \hline
 
  \multirow{2}{*}{Right Hippocampus} & Dice & 0.732 & \textbf{0.759} \\
 \cline{2-4} & HD & 54.553 & \textbf{29.878} \\ 
 \hline
\end{tabular}
\end{center}

\caption{Performance of segmentation, evaluated on both Dice coefficient (Dice) and Hausdorff distance (HD).}
\label{lcn}
\end{table}
\vspace{-5mm}

The Dice coefficient and Hausdorff distance in the tables are both metrics to evaluate the similarity between a segmentation result and its ground truth label map. A higher Dice coefficient and a lower Hausdorff distance both means greater similarity. It's shown that our approach achieves better results than 3D U-Net in terms of both Dice coefficient and Hausdorff distance. In the visual results, it is shown that our approach, compared with 3D U-Net, captures the intricate shape details better. In both examples in Figure \ref{segresult}, 3D U-Net cannot segment the sharp part in the lower part of a caudate nucleus while our method is able to. \par
Since all experiment settings except loss function are the same for 3D U-Net and our method, the better performance of our method is due to the incorporation of shape information. Concretely, the shape loss measures the difference in shape signature, while shape signature extracted by a network trained to minimize the difference in shape signature between two \textit{affine pairs} of the same subject. Therefore, when the difference in shape signature is used as a part of segmentation network's loss function, it naturally guides the segmentation network to produce segmentation results that comply with the shapes they should have, thus having better results both quantitatively and visually. 

\section{Conclusion}
We present a novel approach that incorporates shape information into the task of 3D medical image segmentation, by training an shape-learning network that learns the shape signature of the target to be segmented. We run experiments on the public LPBA40 dataset on the brain structure of caudate nucleus and hippocampus. Experimental results show that our approach leads to better results than 3D U-Net in terms of both Dice coefficient and Hausdorff distance.

\end{document}